\documentclass[12pt]{article}
\usepackage[margin=1in]{geometry}
\usepackage[T1]{fontenc}
\usepackage[utf8]{inputenc}
\usepackage{amsmath,amssymb,graphicx,booktabs,array}
\usepackage{cite}
\usepackage[hidelinks]{hyperref}
\usepackage{caption}
\usepackage{placeins}

\title{Accelerating Fractal Image Compression Using Search Pruning and Fractal Dimension Gating}
\author{Nastaran Pourshab\\University of Cincinnati\\2600 Clifton Ave, Cincinnati, OH 45221, USA\\\texttt{pourshnn@mail.uc.edu}
\and Mohsen Bagheritabar\\University of Illinois Chicago\\Chicago, Illinois 60607, USA\\\texttt{Mbagh@uic.edu}}
\date{}
\begin{document}
\maketitle
\begin{abstract}
Fractal image compression (FIC) can achieve
high compression ratios and fast decoding, but its encoding stage
requires an expensive search for matching image blocks. This paper
evaluates two implementation strategies for reducing this cost: isometry
pruning, and fractal-dimension-based search gating. The gating method
uses a box-counting estimate of range-block complexity to determine
whether the domain-pool search should be performed. Experiments on the
Monkey, Einstein, and Cameraman grayscale images examine two
domain/range block-size configurations. The results show that isometry
pruning substantially reduces total execution time. Combining isometry
pruning with box-counting gating further reduces the number of searched
blocks and the bitrate, although the resulting reduction in PSNR depends
on the selected thresholds. These findings characterize the trade-off
among execution time, bitrate, and reconstruction quality.

\end{abstract}
\section{Introduction}\label{introduction}

Fractal image compression (FIC) is a lossy compression technique that
exploits self-similarity within an image \cite{ref1,ref2,ref3}. Its advantages
include high compression ratios, competitive reconstruction quality, and
fast decoding, whereas its principal limitation is the computational
cost of encoding \cite{ref4,ref5}. Garg et al. \cite{ref6} compared FIC with the
Joint Photographic Experts Group (JPEG) standard, vector quantization
(VQ), and wavelet-based compression. Their comparison identifies
compression efficiency as an advantage of FIC and its long encoding time
as a major limitation.

During FIC encoding, the image is partitioned into nonoverlapping range
blocks and a pool of larger domain blocks. In the implementation studied
here, each domain block has twice the side length of a range block. Each
domain block is reduced to the range-block size and transformed using
rotations and reflections. For every range block, the encoder searches
for the transformed domain block that minimizes the squared
reconstruction error. The selected domain-block coordinates and
transformation parameters are stored and applied iteratively during
decoding \cite{ref1,ref5,ref7}.

The search for similar patterns between domain and range blocks is the
main computational bottleneck in FIC, motivating several search-space
optimization methods \cite{ref3,ref4,ref5,ref8}. Al-Bundi et al. \cite{ref3} reviewed the
harmony search algorithm, particle swarm optimization, the crowding
optimization method, and genetic algorithms. These methods reduce the
number of candidate matches evaluated during encoding, but their
integration can add algorithmic complexity. The present work instead
examines a relatively simple box-counting criterion that can be inserted
before the conventional domain-pool search.

This study modifies an open-source FIC implementation \cite{ref7} in two
stages: pruning the set of evaluated isometries and applying
fractal-dimension-based gating before domain-block matching. Previous
studies have used fractal dimension or differential box counting to
accelerate FIC encoding \cite{ref9,ref10,ref11,ref12}. For example, Al-Saidi and Ali
\cite{ref12} used differential box-counting dimension as a matching feature
for range blocks and domain-block pools with eight-orientation. In the
method proposed here, two thresholds, \(T_{1}\) and \(T_{2}\), determine
whether a range block enters the domain-matching stage. \(T_{1}\) is the
intensity threshold used to determine which boxes are counted when
estimating the fractal dimension, and \(T_{2}\) is the complexity
threshold used to determine whether the estimated block complexity is
high enough to justify the search. The experiments evaluate the effects
of these thresholds and their combination with isometry pruning on
bitrate, PSNR, total execution time, and the number of stored
transformations.

\section{Methods}\label{methods}

\subsection{Evaluation Metrics}\label{evaluation-metrics}

Performance was evaluated using compression ratio (CR), bitrate in bits
per pixel (bpp), total execution time, root mean square error (RMSE)
\cite{ref13}, and peak signal-to-noise ratio (PSNR). For 8-bit grayscale
images, PSNR was calculated as \(20{\log}_{10}(255/RMSE)\).
Rate--distortion performance was evaluated by plotting PSNR against bpp.
The compressed representation size included the bits required for the
domain-block coordinates, rotation, reflection, contrast, and brightness
parameters.

\subsection{Isometry Pruning}\label{isometry-pruning}

The baseline encoder evaluates eight isometries for each domain block:
four rotations, each with and without reflection. The pruned
configuration removes reflection and retains rotations of 0 and 90
degrees. It therefore evaluates two orientations instead of eight for
every candidate domain block. This change reduces both the number of
comparisons and the number of bits required to represent the selected
isometry.

\subsection{Fractal Dimension Based Search
Gating}\label{fractal-dimension-based-search-gating}

The box-counting method provides a preprocessing criterion for deciding
whether a range block should undergo the computationally expensive
domain-pool search. Each range block is divided into boxes at multiple
scales, r. At each scale, a box is considered active when its mean pixel
intensity exceeds the intensity threshold \(T_{1}\). The number of
active boxes, N(r), is then counted. The range block\textquotesingle s
fractal dimension is estimated from the slope of the linear relationship
between log N(r) and log(1/r) \cite{ref14,ref15,ref16}:

\[D_{i} = \frac{\Delta \log N(r)}{\Delta \log(1/r)}.\]

The estimated dimension, D, is compared with the complexity threshold
\(T_{2}\). If \(D>T_2\), the range block is classified
as complex, and the conventional FIC search identifies its best matching
transformed domain block. If \(D\leq T_2\), the block is
classified as simple and the domain-pool search is skipped. Thus,
\(T_{1}\ \)determines which boxes contribute to the dimension estimate,
whereas \(T_{2}\ \)determines whether a range block enters the matching
stage. Increasing \(T_{2}\ \)generally causes more blocks to bypass the
search. This can reduce the number of stored transformations and the
bitrate, but it may also increase reconstruction error.

\begin{figure}[htbp]
\hrule\medskip
\textbf{Algorithm 1: Fractal-dimension-based range-block selection}

\smallskip
\textbf{Input:} Image $I$; range-block size $R$; domain-block size $D$;
intensity threshold $T_1$; complexity threshold $T_2$.

\textbf{Output:} Encoded FIC parameters.
\begin{enumerate}
\item Partition $I$ into range blocks.
\item Construct the domain-block pool.
\item For each range block $R_i$:
\begin{enumerate}
\item Divide $R_i$ into boxes at multiple scales $r$.
\item At each scale, count boxes whose mean intensity exceeds $T_1$.
\item Estimate $D_i$ from the slope of $\log N(r)$ versus $\log(1/r)$.
\item If $D_i>T_2$: search the transformed domain-block pool; select the domain block and affine parameters that minimize reconstruction error; store the resulting transformation.
Otherwise: skip the domain-pool search.
\end{enumerate}
\item Return all stored parameters.
\end{enumerate}
\hrule
\end{figure}

\section{Results}\label{results}

\subsection{Baseline Results}\label{baseline-results}

The baseline implementation was evaluated in the Kaggle environment
using three grayscale images: Monkey (256 x 256 pixels), Einstein (512 x
512 pixels), and Cameraman (256 x 256 pixels). Table~\ref{tab:results} reports bitrate,
PSNR, total execution time, and the number of searched blocks for
domain/range block sizes of 32/16 and 16/8.

\subsection{Effect of Isometry
Pruning}\label{effect-of-isometry-pruning}

Isometry pruning increased the compression ratio from 81.92 to 89.04 for
the Monkey image and from 75.85 to 81.92 for the Einstein image. It also
reduced total execution time because each domain block was evaluated
under two orientations instead of eight.

\subsection{Effect of Fractal Dimension Based Search
Gating}\label{effect-of-fractal-dimension-based-search-gating}

The threshold values were varied to examine how fractal-dimension-based
gating affects compression performance. Table~\ref{tab:results} summarizes
representative baseline, gating, isometry-pruning, and combined results
for the three test images and both block-size configurations.
\(T_{1}\) is the intensity threshold used when counting active boxes. A
box contributes to N(r) when its mean pixel intensity exceeds \(T_{1}\).
\(T_{2}\ \)is the complexity threshold applied to the estimated fractal
dimension. If a range block's estimated fractal dimension exceeds
\(T_{2}\), the encoder searches the domain-block pool for its best match
and stores the corresponding transformation parameters. Otherwise, the
domain-block search is skipped, and no transformation is stored for that
range block. Increasing \(T_{2}\) generally reduces the number of range
blocks that undergo domain-block matching and, consequently, the number
of stored transformations. However, it may also reduce reconstruction
quality. The combined configuration applies box-counting gating and
isometry pruning. Isometry pruning accounts for most of the decrease in
total execution time because it reduces the number of orientations
evaluated for every searched domain block. Box-counting gating further
reduces the number of range blocks that enter the matching stage, but
computing the fractal dimension adds preprocessing overhead.
Consequently, gating alone can take longer than the baseline when too
few searches are skipped, whereas the combined method remains faster
because each retained search is substantially less expensive. Table~\ref{tab:results}
also shows that reductions in bitrate and total execution time can be
accompanied by lower PSNR.

\begin{table}[p]
\centering
\caption{Rate--distortion and total-execution-time results for the Monkey, Einstein, and Cameraman images using domain/range block sizes of 32/16 and 16/8.}
\label{tab:results}
\small
\setlength{\tabcolsep}{3pt}
\resizebox{\textwidth}{!}{%
\begin{tabular}{llcrrrrrr}
\toprule
Image & Configuration & \shortstack{Domain/\\range} & $T_1$ & $T_2$ & bpp & \shortstack{PSNR\\(dB)} & \shortstack{Total execution\\time (s)} & \shortstack{Searched\\blocks}\\
\midrule

Monkey & Baseline & 32/16 & - & - & 0.0977 & 25.95 &
4.497 & 256 \\
& Box counting & 32/16 & 50 & 1.6 & 0.0889 & 24.39 & 8.631 & 233 \\
& Isometry pruning & 32/16 & - & - & 0.0898 & 25.38 & 2.618 & 256 \\
& Combined & 32/16 & 50 & 1.6 & 0.082 & 23.89 & 2.758 & 233 \\
& Baseline & 16/8 & - & - & 0.4219 & 28.99 & 58.967 & 1024 \\
& Box counting & 16/8 & 50 & 1.6 & 0.3398 & 22.13 & 88.598 & 825 \\
& Isometry pruning & 16/8 & - & - & 0.3906 & 28.74 & 30.065 & 1024 \\
& Combined & 16/8 & 50 & 1.6 & 0.3147 & 21.89 & 23.315 & 825 \\
\midrule Einstein & Baseline & 32/16 & - & - & 0.1054 & 24.10 &
70.80 & 1024 \\
& Box counting & 32/16 & 20 & 2 & 0.0981 & 22.50 & 132.358 & 952 \\
& Isometry pruning & 32/16 & - & - & 0.0977 & 23.77 & 42.83 & 1024 \\
& Combined & 32/16 & 20 & 2 & 0.0908 & 22.38 & 34.535 & 952 \\
& Baseline & 16/8 & - & - & 0.4531 & 28.51 & 937.991 & 4096 \\
& Box counting & 16/8 & 20 & 2 & 0.4169 & 23.51 & 1569.158 & 3769 \\
& Isometry pruning & 16/8 & - & - & 0.4219 & 28.22 & 463.696 & 4096 \\
& Combined & 16/8 & 20 & 2 & 0.3882 & 23.46 & 404.45 & 3769 \\
\midrule Cameraman & Baseline & 32/16 & - & - & 0.0977 & 22.42 &
6.148 & 256 \\
& Box counting & 32/16 & 20 & 2 & 0.0893 & 20.08 & 8.964 & 234 \\
& Isometry pruning & 32/16 & - & - & 0.0898 & 22.03 & 2.765 & 256 \\
& Combined & 32/16 & 20 & 2 & 0.0821 & 20.15 & 2.969 & 234 \\
& Baseline & 16/8 & - & - & 0.4219 & 25.99 & 62.238 & 1024 \\
& Box counting & 16/8 & 20 & 2 & 0.3749 & 20.21 & 100.285 & 910 \\
& Isometry pruning & 16/8 & - & - & 0.391 & 25.48 & 32.464 & 1024 \\
& Combined & 16/8 & 20 & 2 & 0.3471 & 19.16 & 25.637 & 910 \\
\bottomrule
\end{tabular}}
\end{table}

\begin{figure}[p]
\centering
\includegraphics[width=.72\textwidth]{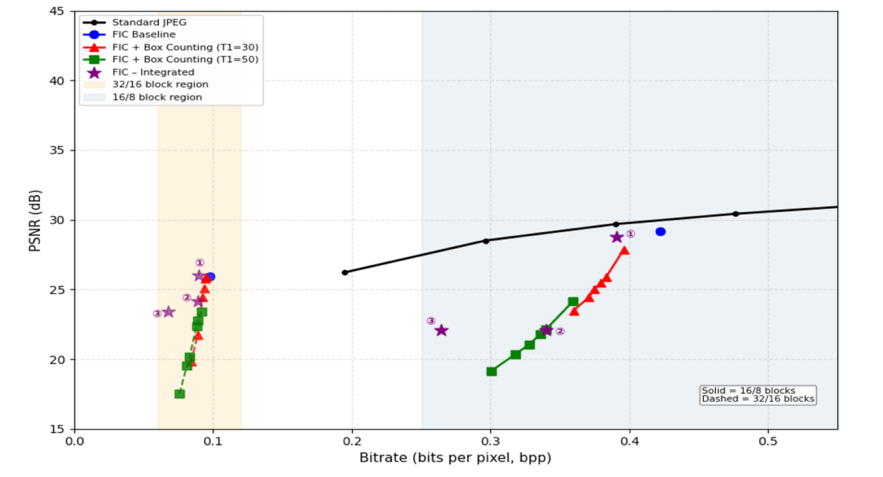}\\[5pt]
\includegraphics[width=.72\textwidth]{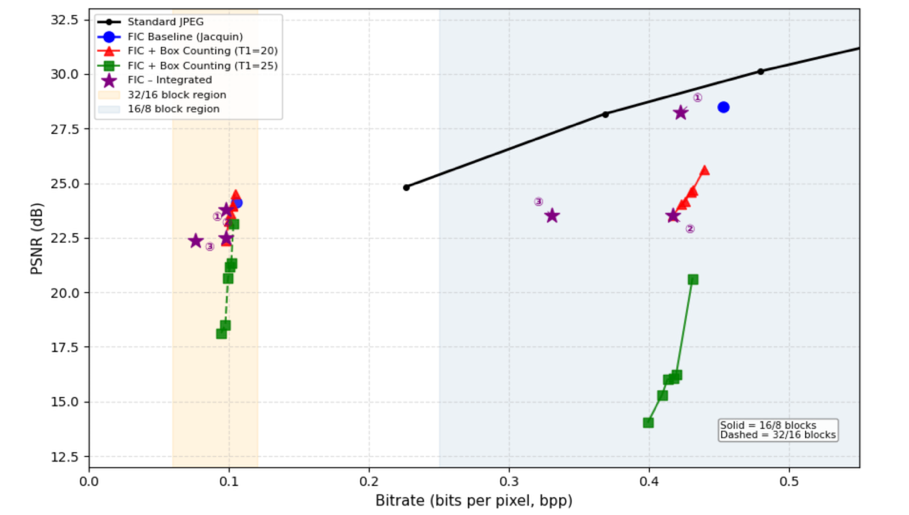}\\[5pt]
\includegraphics[width=.72\textwidth]{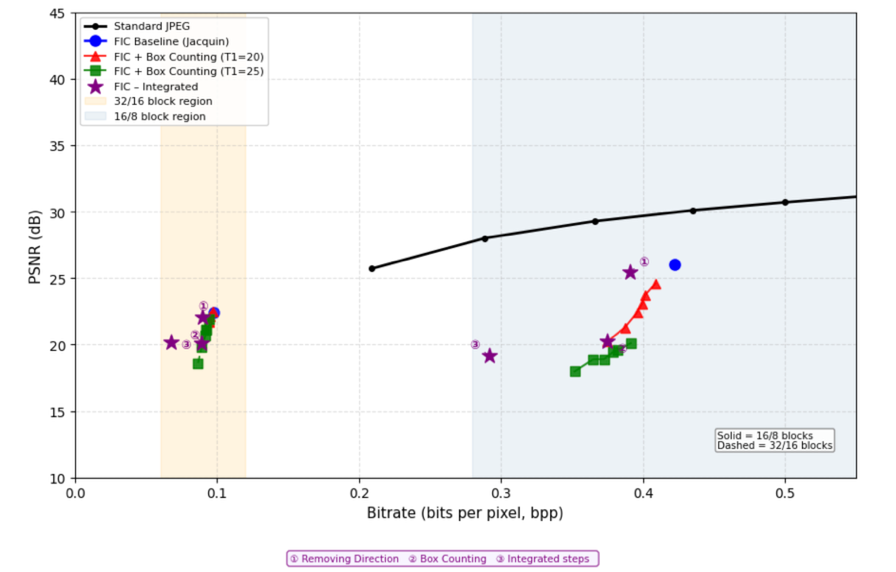}
\caption{Rate--distortion performance for the Monkey, Einstein, and Cameraman images, including the evaluated FIC configurations and JPEG reference curves.}
\label{fig:rate-distortion}
\end{figure}

Figure~\ref{fig:rate-distortion} presents the rate--distortion performance for the Monkey,
Einstein, and Cameraman images. In these plots, the bitrate is expressed
in bits per pixel (bpp), while the reconstruction quality is measured
using peak signal-to-noise ratio (PSNR). A point located toward the
upper-left region is generally preferable because it represents higher
reconstruction quality at a lower bitrate. For all three images,
changing the box-counting thresholds produces a trade-off between
bitrate and reconstruction quality. Increasing the gating threshold
causes more range blocks to bypass the domain-matching search.
Consequently, fewer transformations are stored and the bitrate
decreases; however, the reconstruction error increases and the PSNR
generally decreases. The 32/16 block configuration operates in a
lower-bitrate region than the 16/8 configuration because it requires
fewer transformations. In contrast, the smaller 16/8 blocks generally
preserve more local image detail but require a higher bitrate.

The integrated method combines fractal-dimension-based gating with
isometry pruning. Its operating points demonstrate that the proposed
optimizations can reduce the bitrate and the number of evaluated
transformations, although aggressive pruning or gating can reduce
reconstruction quality. Therefore, the threshold values should be
selected to balance compression efficiency and image quality rather than
to minimize bitrate alone. The results also vary among the three test
images, indicating that the rate--distortion performance depends on
image content as well as block size and threshold values. The JPEG
curves are included as reference points; they generally provide higher
PSNR at comparable or higher bitrates, while the proposed FIC
configurations explore lower-bitrate operating regions and reductions in
the domain-search workload.

\begin{figure}[htbp]
\centering
\includegraphics[width=.60\textwidth]{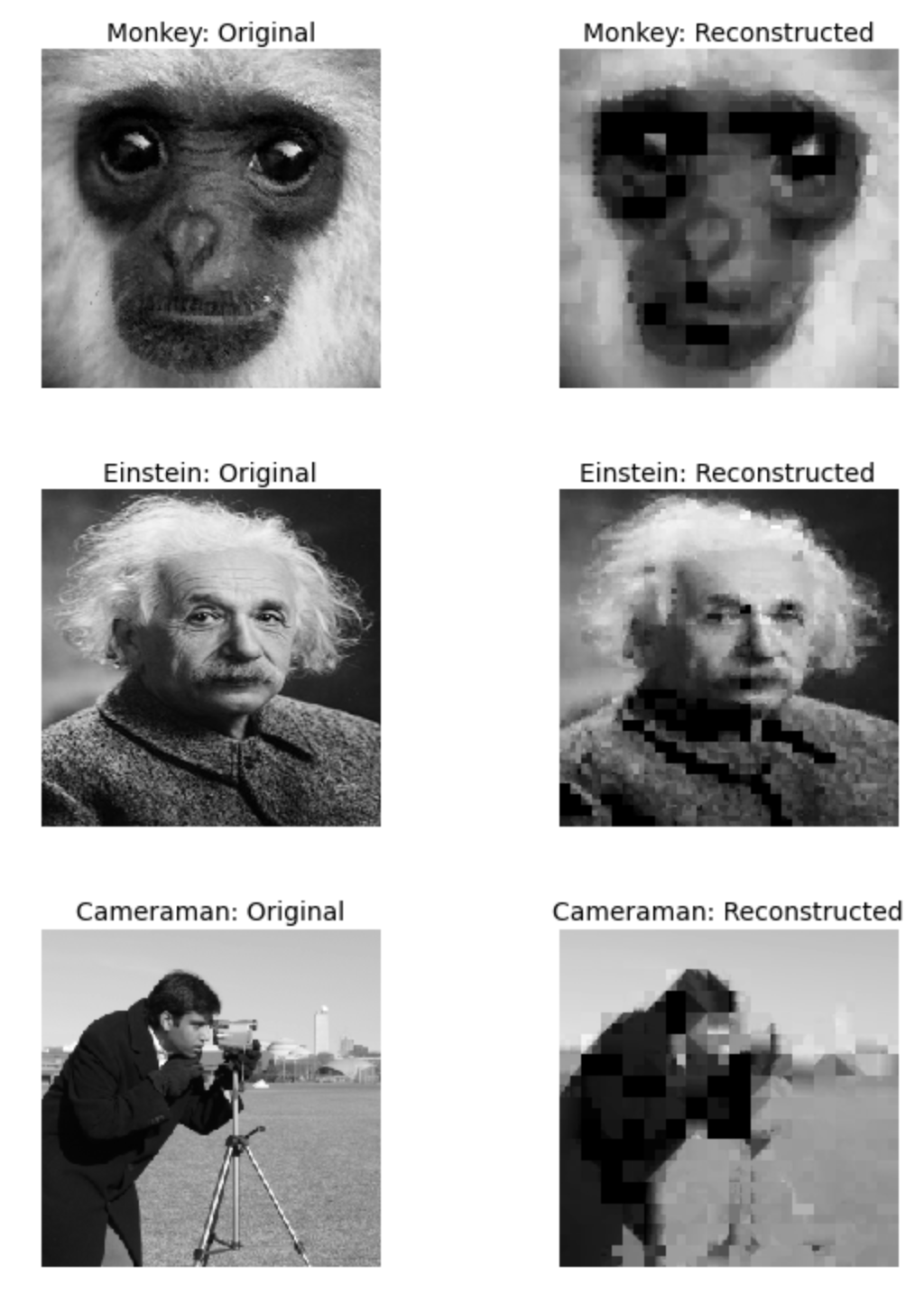}
\caption{Original and reconstructed test images obtained using
the proposed combined method: Monkey (top row; (\(T_{1}\)=50),
(\(T_{2}\)=1.6)) and Einstein and Cameraman (middle and bottom rows,
respectively; (\(T_{1}\)=20), (\(T_{2}\)=2)). Domain/range block sizes
of 32/16 pixels were used for all three images.}
\label{fig:reconstructions}
\end{figure}

Figure~\ref{fig:reconstructions} presents the original and reconstructed versions of the Monkey,
Einstein, and Cameraman test images obtained using the proposed method.
In all three cases, the reconstructed images preserve the principal
objects and overall spatial structure of the originals. However, some
fine details and smooth intensity variations are lost, and block
artifacts are visible, particularly in textured regions. These effects
arise from the block-based approximation used in fractal image
compression. The Cameraman reconstruction exhibits more visible block
artifacts and greater loss of detail than the Monkey and Einstein
reconstructions. The result suggests that the effectiveness of the
proposed method depends on image content and selected threshold
configuration.

\section{Conclusion}\label{conclusion}

This paper evaluated isometry pruning, and fractal-dimension-based
search gating in an open-source FIC encoder. Isometry pruning produced
the largest reduction in execution time by decreasing the number of
orientations evaluated for each candidate block. Box-counting gating
reduced the number of searched blocks and the bitrate, but its
preprocessing overhead sometimes increased encoding time when used
alone. Combining gating with isometry pruning reduced execution time
relative to the baseline for all reported test cases, although PSNR
decreased as more blocks bypassed the search. The thresholds should
therefore be selected according to the required trade-off among bitrate,
encoding time, and reconstruction quality.

\FloatBarrier

\end{document}